\documentclass[aps,amssymb,amsmath,
twocolumn,showpacs,superscriptaddress,
groupedaddress,ifmbe]{revtex4-1}

\usepackage{graphicx}
\usepackage{dcolumn}
\usepackage{bm}
\usepackage[table,xcdraw]{xcolor}
\usepackage{nth}
\usepackage[ruled,vlined]{algorithm2e}
\usepackage{framed}
\usepackage{listings}
\usepackage[titletoc,title]{appendix}
\usepackage{tikz}
\usepackage{lipsum}
\usepackage{hyperref}
\usepackage{multirow}
\usepackage{nicefrac}
\usepackage[table]{xcolor}

\hypersetup{
  colorlinks,
  citecolor=red,
  linkcolor=blue,
  urlcolor=blue}

\begin{document}
\title{Relativistic exponential$-$type spinor orbitals\\ and their use in many$-$electron Dirac equation solution}
\author{A. Ba{\u g}c{\i}}
\email{abagci@pau.edu.tr}
\affiliation{Department of Physics, Faculty of Sciences, Pamukkale University, {\c C}amlaraltı, K{\i}n{\i}kl{\i} Campus, Denizli, Turkey}

\begin{abstract}
Dirac$-$Coulomb type differential equation and its solution relativistic exponential$-$type spinor orbitals are introduced. They provide a revised form for operator invariants, namely Dirac invariants, simplifying the treatment of the angular components in calculation of many$-$electron systems. The relativistic Coulomb energy is determined by employing a spectral solution to Poisson's equation for the one$-$electron potential, which is expressed in terms of radial functions involving incomplete gamma functions. The computation for incomplete gamma functions posses challenges due to slow convergence rate associated with their series representation. Such difficulties are eliminated through use of the bi$-$directional method along with hyper$-$radial functions. A new formulation for relativistic auxiliary functions that improve the efficiency in Coulomb energy calculations is presented. These formulations also contribute to inquiring into orthogonal expansions for solutions to Poisson's equation using complete orthonormal sets of exponential orbitals with non$-$integer principal quantum numbers. They may provide a meaningful alternative series representations.
\begin{description}
\item[Keywords]
Dirac equation, Exponential$-$type spinor orbitals, Hyper$-$radial functions
\item[PACS numbers]
... .
\end{description}
\end{abstract}
\maketitle
\section{Introduction} \label{introduction}
The hydrogenic Dirac equation \cite{1_Dirac_1928, 2_Dirac_1930} solutions \cite{3_Quiney_1989, 4_Grant_2000, 5_Bagci_2016, 6_Bagci_2020} reveal that basis functions used in studying many$-$electron systems with the linear combination of atomic spinors method (LCAS) \cite{7_Roothaan_1951, 8_Kim_1967, 9_Malli_1975, 10_Quiney_1987, 11_Yanai_2001} are expected to possess non$-$integer quantum numbers. The Dirac equation for quantum mechanical characterization of an electron moving through the Coulomb potential around a fixed$-$point$-$like nucleus of charge $Ze$, where $e$ is the proton charge and $Z$ is the atomic number, is given as,
\begin{align}\label{eq:1}
\hat{H}_{D}\Psi=E\Psi,
\end{align}
where,
\begin{align}\label{eq:2}
\hat{H}_{\mathit{D}}=c(\vec{\alpha}.\hat{\vec{p}})+m_{0}c^{2}\beta-\frac{Ze^2}{r}
\end{align}
is the one-electron Dirac operator (in atomic units (a.u.); $\hslash=1$, $m_{0}=1$ and ${e^2}/{4\pi\epsilon_{0}}=1$).\\
Following the standard terminology for positive energy solutions, the two$-$component form of the four$-$component electron spinor wave function can be expressed as: \cite{12_Foldy_1950},
\begin{align}\label{eq:3}
\Psi_{p^{*}}\left(\vec{r}\right)=
\begin{pmatrix}
\psi^{L}_{p_{1}^{*}}\left(\vec{r}\right)
\\
\psi^{S}_{p_{2}^{*}}\left(\vec{r}\right)
\end{pmatrix}.
\end{align}
with labels $L,S$, used to denote the terms \textit{large} and \textit{small}, respectively. $p$, $p_{i}$ are used to represent sets of quantum numbers. In two$-$component form of solution to the Dirac equation $p$, $p_{i}$ are defined as, $p=p_{1}=\left\lbrace n^{*},\kappa, m \right\rbrace$, $p_{2}=\left\lbrace n^{*}, -\kappa, m \right\rbrace$. Here, $n^{*}$, $n^{*} \in {\mathbb{R}}/{\mathbb{N}}$ are the principal quantum numbers, $\kappa$. $\kappa=\left\lbrace \pm 1, \pm 2, ...\right\rbrace$ are the secondary total angular momentum quantum numbers and $m$ are the magnetic quantum numbers.\\
In the non$-$relativistic limit, the lower components of Eq. (\ref{eq:3}) go to zero, while the upper components become a solution to the corresponding non$-$relativistic equation, namely the Schr{\"o}dinger equation \cite{13_Greiner_1997}. A recent development has seen the derivation of complete and orthonormal wave$-$functions with non$-$integer quantum numbers, applicable to solving Schr{\"o}dinger$-$like equations \cite{14_Bagci_2023}. 
\begin{itemize}
\item{This study primarly focus on spinor wave$-$functions as solutions to Dirac$-$type equations. Their non$-$relativistic limit directly corresponds these newly derived wave$-$functions.}
\end{itemize}

In terms of many$-$electron systems, aforementioned generalization for one$-$electron Hamiltonian eigen$-$functions also requires revisiting the application of the Poisson equation to the quantum mechanical Coulomb problem.\\
One can express the Coulomb energy related with charge density $\rho\left(\vec{r}\right)$ as \cite{15_Manby_2001},
\begin{align}\label{eq:4}
E\left[\rho\right]=\int\int \frac{\rho\left(\vec{r}_{1}\right)\rho\left(\vec{r}_{2}\right)}{\vert \vec{r}_{1}-\vec{r}_{2} \vert}d\vec{r}_{1}d\vec{r}_{2}
\end{align}
Applying Poisson's equation transforms the Eq. (\ref{eq:3}) into one$-$electron kinetic energy$-$type integral:
\begin{multline}\label{eq:5}
E\left[\rho\right]=
-\frac{1}{4\pi}\int V\left(\vec{r}\right) \Delta^{2}_{\vec{r}} V^{*}\left(\vec{r}\right)d\vec{r}\\
=\int V\left(\vec{r}\right)\rho\left(\vec{r}\right)d\vec{r}
\end{multline}
Eq. (\ref{eq:5}) is obtained by applying Green's theorem, taking into account both the Laplacian's Green's function and the charge density as described by the Poisson equation. Here, $V$ is the one$-$center potential and $V^{*}$ is the complex conjugate of $V$.\\
Within the relativistic framework, calculating the Coulomb energy through one$-$center potential expansion using spectral forms \cite{16_Weatherford_2005, 17_Weatherford_2006} demands employing a new set of basis functions characterized by non$-$integer principal quantum numbers \cite{18_Bagci_2015},
\begin{align}\label{eq:6}
V\left(\vec{r}\right)=\sum_{n^{*}lm}F_{n^{*}lm}\left(\zeta, \vec{r}\right)\mathcal{C}_{n^{*}lm}
\end{align}
The spectral forms are separated in spherical coordinates as,
\begin{align}\label{eq:7}
F_{n^{*}lm}\left(\zeta, \vec{r}\right)=
\mathcal{N}_{n^{*}}\left(\zeta\right)f_{n^{*}l}\left(\zeta r\right)
Y_{lm}\left(\theta, \varphi\right)
\end{align}
$Y_{lm}$ are complex or real spherical harmonics \cite{19_Rose_1957}. The radial functions are given as,
\begin{multline}\label{eq:8}
f_{n^{*}l}\left(x\right)
=\Gamma\left(n^{*}+l+1\right)\frac{1}{x^{l+1}}
\biggl\{
P\left[n^{*}+l+1,x\right]
\Bigg. 
\\
\Bigg.
+\frac{x^{2l+1}}{\left(n^{*}-l\right)}Q\left[n^{*}-l,x\right].
\biggl\}
\end{multline}
$P\left[a,x\right]$, $Q\left[a,x\right]$ are regularized gamma functions with,
\begin{align}\label{eq:9}
P\left[a,x\right]&=\frac{\gamma\left[a,x\right]}{\Gamma\left(a\right)},
&
Q\left[a,x\right]&=\frac{\Gamma\left[a,x\right]}{\Gamma\left(a\right)},
\end{align}
$\gamma\left[a,x\right]$, $\Gamma\left[a,x\right]$ are incomplete and complementary incomplete gamma function, respectively \cite{20_Abramowitz_1974}. The expression for the radial functions remain consistent with that in \cite{16_Weatherford_2005}. However, the gamma functions now involve parameters with non$-$integer values, which complicates the computation of both the radial functions and the associated integrals.\\
The atomic and molecular two$-$electron Coulomb energy have recently been expressed by the authors in terms of new hyper$-$radial functions \cite{21_Bagci_2024} and relativistic molecular auxiliary functions \cite{18_Bagci_2015, 22_Bagci_2018, 23_Bagci_2018, 24_Bagci_2020, 25_Bagci_2022}. These functions have been obtained through bi$-$directional method \cite{21_Bagci_2024} and revisiting procedure of solving Poisson's equation for two$-$center case using spectral forms, respectively. Symmetry features of the Coulomb energy, as defined in \cite{22_Bagci_2018} by a criterion, are used to obviate the need for immediate expansion of incomplete gamma functions. Direct computation of the Eq. (\ref{eq:8}) however, depends on accurate calculation of incomplete gamma functions. Additionally, these functions are encountered when addressing the quantum electrodynamics effects by Dirac$-$Coulomb$-$Breit equation \cite{26_Jentschura_2002, 27_Jeszenszki_2022} (see also references therein). The incomplete gamma functions were investigated by numerous authors from the mathematical point of view \cite{28_Temme_1994, 29_Gautschi_1999, 30_Ferreira_2005, 31_Amore_2005}. They are yet, still subject to some research because various domains of convergence, contingent upon parameters, can be constructed for any representation (recurrence relationships, continued fractions, infinite series expansion formulas, or asymptotic methods) of them \cite{32_Gil_2012, 33_Backeljauw_2014, 34_Bujanda_2018, 35_Abergel_2020}. 
\begin{itemize}
\item{The second aspect of this study is to demonstrate how the symmetry properties arising in physical systems can be utilized to provide insights into a longstanding problem in mathematics. We consider the bi$-$directional method, allows benefit form the symmetry features of Coulomb energy to calculate the most basic higher transcendental functions, incomplete beta and gamma functions.}
\end{itemize}

Consider the following expansion for a function $x^{n^{*}}e^{\zeta r}$ \cite{36_Weniger_2008},
\begin{multline}\label{eq:10}
x^{n^{*}}e^{\zeta r}
=\left(1-\zeta\right)^{-n^{\prime *}-n^{*}-1}
\frac{\Gamma\left(n^{\prime *}+n^{*}+1\right)}{\Gamma\left(n^{\prime *}+1\right)}
\\
\times
\sum_{k=0}^{\infty} {}_{2}F_{1} \left[
-k, n^{\prime *}+n^{*}+1; n^{\prime *}+1;
\frac{1}{1-\zeta}
\right]
\\
\times L_{k}^{n^{\prime *}}\left(x\right)
\end{multline}
$n^{*} \in \mathbb{R}/ \mathbb{N}$, $\mathbb{R}e\left(n^{*}+n^{\prime *}\right)>-1$, $\zeta \in \left(-\infty, 1/2 \right)$.
${}_{2}F_{1}\left[a,b;c;x\right]$ are Gauss hyper$-$geometric functions \cite{37_Magnus_1966}. $L_{q-p}^{p}\left(x\right)$ are Laguerre polynomials. $\Gamma\left(n^{*}\right)$ is the gamma function. Eq. (\ref{eq:10}) plays an essential role in understanding the one-center exponential$-$type electron wave$-$function expansion. $x^{n^{*}}e^{\zeta r}$ is however, analytic at the origin only if $n^{*}=n$, $n \in \mathbb{N}$. When $\zeta=0$, the series simplifies to a single, closed$-$form expression due to its termination \cite{36_Weniger_2008}.\\
In case that $n^{*} \in \mathbb{R}/ \mathbb{N}$, the Eq. (\ref{eq:10}) simplifies to,
\begin{align}\label{eq:11}
x^{n^{*}}=\frac{\Gamma\left(n^{*}+n^{\prime *}+1\right)}{\Gamma\left(n^{\prime *}\right)}\sum_{k=0}^{\infty}\frac{\left(-n^{*}\right)_{k}}{\left(n^{\prime *}+1\right)_{k}}L_{k}^{n^{\prime *}}\left(x\right),
\end{align}
$\left(n^{*}\right)_{k}=\Gamma\left(n^{*}+k\right) / \Gamma\left(n^{*}\right)$ is the Pochhammer symbol. After substituting the explicit expression of generalized Laguerre polynomials into the equation and rearranging order of summation, the expression is found to involve generalized ${}_{1}F_{0}\left[k-n^{*};;x\right]$ hyper$-$geometric functions with $x=1$, converging only if $\vert x \vert < 1$.
\begin{itemize}
\item{The third aspect of this study is analysis of non$-$existence of expansions in terms of orthogonal polynomials for general power functions, a nontrivial research area. This analysis is conducted through the aforementioned approximations presented by the authors.}
\end{itemize}
\section{Relativistic Exponential$-$type Spinors Orbitals}
Newtonian generalization for radial parts of non$-$relativistic exponential$-$type orbitals to non$-$integer quantum numbers are given as \cite{14_Bagci_2023},
\begin{align}\label{eq:12}
\Psi_{n^{*}lm}^{\alpha \varepsilon}\left(\vec{r},\zeta\right)
=R_{n^{*}l}^{\alpha \epsilon}\left(r,\zeta\right)
Y_{lm}\left(\theta,\varphi\right),
\end{align}
Here, the radial parts $R_{n^{*}l}^{\alpha \epsilon}$ are dedined as,
\begin{align}\label{eq:13}
R_{n^{*}l}^{\alpha \epsilon}\left(r,\zeta\right)
=N_{n^{*}l}^{\alpha \varepsilon}\left(\zeta\right)
\left(2\zeta r\right)^{l+\varepsilon-1}
L_{n^{*}-l-\varepsilon}^{2l+2\varepsilon-\alpha}\left(2\zeta r\right)
\end{align}
where, $\varepsilon \neq 0$, $\varepsilon \leq 1$, $\zeta>0$ and $\alpha>-3$. $N_{n^{*}l}$ are normalization constants,
\begin{align}\label{eq:14}
N_{n^{*}l}^{\alpha \varepsilon}\left(\zeta\right)
\left[
\frac{\left(2\zeta\right)^{3}\Gamma\left(n^{*}-\varepsilon-l+1\right)}
{\left(2n^{*}\right)^{\alpha}\Gamma\left(n+\varepsilon+l+1-\alpha\right)}
\right]^{1/2}
\end{align}
The Eq. (\ref{eq:13}) was established via following orthogonality relationship for the generalized Laguerre polynomials,
\begin{multline}\label{eq:15}
\int_{0}^{\infty} \left(2\zeta r\right)^{p^{*}}
e^{-2\zeta r}
L_{q^{*}-p^{*}}^{p^{*}}\left(2\zeta r\right)
L_{q^{\prime *}-p^{\prime *}}^{p^{*}}\left(2\zeta r\right)dr
\\
=\frac{\Gamma\left(q^{*}+1\right)}{\Gamma\left(q^{*}-p^{*}+1\right)}
\delta_{q^{*}-p^{*},q^{\prime *}-p^{\prime *}},
\end{multline}
and methodology of fractional calculus. For a comprehensive treatment of fractional calculus refer to \cite{38_Oldham_1974, 39_Kilbas_2006, 40_Kochubei_2019}.\\
Recently postulated variants of quantum numbers, based on the definition of Laguerre polynomials for fractional values, are first generalized to the relativistic case. This is achieved by considering the non$-$relativistic limit of the Dirac$-$Coulomb equation, where the solutions should satisfy the \textit{generalized} Coulomb$-$Sturmians ($\alpha=1$ in the Eq. (\ref{eq:13})). Here, \textit{generalized} refers to Coulomb$-$Sturmians with quantum numbers that can take non$-$integer values.\\
We start by definition of the radial quantum numbers that caunt the number of nodes or zeros in large components \cite{13_Greiner_1997, 41_Landau_1977}
\begin{align}\label{eq:16}
n^{}_{r}=n^{*}-\vert \kappa^{*} \vert
\end{align}
Power series solution of hydrogenic the Dirac equation leading two more terms \cite{42_Grant_2007},
\begin{align}\label{eq:17}
\gamma^{*}&=\sqrt{{\kappa^{*}}^{2}-\frac{Z^2}{c^2}},
&
N^{*}_{n^{}_{r},\kappa^{*}}&=\sqrt{{n^{}_{r}}^{2}+2n^{}_{r}\gamma^{*}+{\kappa^{*}}^{2}}
\end{align}  

In the non$-$relativistic limit $\left(c\rightarrow \infty\right)$, from Eq. (\ref{eq:13}) we can write,\\
For $\kappa^{*}<0$,
\begin{multline}\label{eq:18}
\kappa^{*}=-\left(l+\varepsilon\right),
\quad
n^{}_{r}=n^{*}-\left(l+\varepsilon\right),
\quad
\gamma^{*}=\vert \kappa^{*} \vert,
\\
N^{*}_{n^{}_{r},\kappa^{*}}=\vert n^{}_{r}+\kappa^{*} \vert
=n^{*}-\left(l+\varepsilon\right)+l+\varepsilon=n^{*} .
\end{multline}
Thus,
\begin{align}\label{eq:19}
\frac{N^{*}_{n^{}_{r},\kappa^{*}}-\kappa^{*}}{n^{}_{r}+2\gamma^{*}}=1
\end{align}
For $\kappa^{*}>0$,
\begin{multline}\label{eq:20}
\kappa^{*}=l+\varepsilon-1,
\quad
n^{}_{r}=n^{*}-l-\varepsilon+1,
\quad
\gamma^{*}=\vert \kappa^{*} \vert,
\\
N^{*}_{n^{}_{r},\kappa^{*}}=\vert n^{}_{r}+\kappa^{*} \vert
=n^{*}-\left(l+\varepsilon-1 \right)+l+\varepsilon-1=n^{*},
\end{multline}
and
\begin{align}\label{eq:21}
\frac{N^{*}_{n^{}_{r},\kappa^{*}}-\kappa^{*}}{n^{}_{r}+2\gamma^{*}}
=\frac{n^{}_{r}}{n^{}_{r}+2l+2\varepsilon-2} .
\end{align}
In line with the established theory of $L-$spinors \cite{3_Quiney_1989, 4_Grant_2000, 42_Grant_2007}, the non$-$relativistic limit of the large components in relativistic exponential$-$type spinors manifests a specific form, represented by:\\
For $\kappa^{*}<0$,
\begin{multline}\label{eq:22}
\lim_{c\to\infty}
{f^{\varepsilon}}^{L}_{n^{}_{r}l^{*}}\left(2\zeta r\right)=
const.\left(2\zeta r\right)^{l+\epsilon}e^{-\zeta r}
\\ \times
\left\lbrace
-L_{n^{}_{r}-1}^{2l+2\varepsilon}\left(2\zeta r\right)
+L_{n^{}_{r}}^{2l+2\varepsilon}\left(2\zeta r\right)
\right\rbrace
\end{multline}
Applying the following relationship for Laguerre polynomials \cite{37_Magnus_1966},
\begin{align}\label{eq:23}
L_{q^{*}-p^{*}}^{p^{*}-1}\left(x\right)
=L_{q^{*}-p^{*}}^{p^{*}}\left(x\right)
-L_{q^{*}-p^{*}-1}^{p^{*}}\left(x\right)
\end{align}
We have,
\begin{multline}\label{eq:24}
\lim_{c\to\infty}
{f^{\varepsilon}}^{L}_{n^{}_{r}l^{*}}\left(2\zeta r\right)=
const.R_{n^{*}l}^{\alpha \epsilon}\left(r,\zeta\right)
\\
=const.\left(2\zeta r\right)^{l+\epsilon}e^{-\zeta r}
L_{n^{}_{r}}^{2l+2\varepsilon-1}\left(2\zeta r\right)
\end{multline}
For $\kappa^{*}>0$,
\begin{multline}\label{eq:25}
\lim_{c\to\infty}
{f^{\varepsilon}}^{L}_{n^{}_{r}l^{*}}\left(2\zeta r\right)=
const.\left(2\zeta r\right)^{l+\varepsilon-1}e^{-\zeta r}
\\
\times \biggl\{
-L_{n^{}_{r}-1}^{2l+2\varepsilon-2}\left(2\zeta r\right)
\biggr.
\\
\biggl.
+\frac{n^{}_{r}}{n^{}_{r}+2l+2\varepsilon-2}
L_{n^{}_{r}}^{2l+2\varepsilon-2}\left(2\zeta r\right)
\biggr\} .
\end{multline}
With regard to a distinct form of Laguerre polynomials \cite{37_Magnus_1966},
\begin{multline}\label{eq:26}
\left(q^{*}+1\right)L_{q^{*}-p^{*}}^{p^{*}}\left(x\right)
-\left(q^{*}-p^{*}+1\right)L_{q^{*}-p^{*}+1}^{p^{*}}\left(x\right)
\\
\times
=x L_{q^{*}-p^{*}}^{p^{*}+1}\left(x\right),
\end{multline}
we obtain an expression identical to Eq. (\ref{eq:24}).\\
The given definitions for quantum numbers, provide accurate description for solution in the non$-$relativistic limit, prompt us to seek for a solution such that it fulfills the following criteria:
\begin{multline}\label{eq:27}
\left\lbrace
{f^{\varepsilon}}^{L}_{n^{}_{r}l^{*}}\left(2\zeta r\right)
+{f^{\varepsilon}}^{S}_{n^{}_{r}l^{*}}\left(2\zeta r\right)
\right\rbrace
\\
=-\left(2\zeta r\right)^{\gamma^{*}}e^{-\zeta r}L_{n^{}_{r}-1}^{2\gamma^{*}}\left(2\zeta r\right)
\end{multline}
\begin{multline}\label{eq:28}
\left\lbrace
{f^{\varepsilon}}^{L}_{n^{}_{r}l^{*}}\left(2\zeta r\right)
-{f^{\varepsilon}}^{S}_{n^{}_{r}l^{*}}\left(2\zeta r\right)
\right\rbrace
\\
=\frac{N^{*}_{n^{}_{r}}-\kappa^{*}}{n^{}_{r}+2\gamma^{*}}
\left(2\zeta r\right)^{\gamma^{*}}e^{-\zeta r}L_{n^{}_{r}}^{2\gamma^{*}}
\left(2\zeta r\right) .
\end{multline}
The radial parts relativistic exponential$-$type spinor orbitals are extracted as,
\begin{multline}\label{eq:29}
{f^{\varepsilon}}^{\beta}_{n^{}_{r}l^{*}}\left(2\zeta r\right)
=\left(2\zeta r\right)^{\gamma^{*}}e^{-\zeta r}
\biggl\{
-\frac{\left(1-\delta_{n^{*}_{r}0}\right)}{2}L_{n^{}_{r}-1}^{2\gamma^{*}}\left(2\zeta r\right)
\biggr.
\\
\biggl.
+\beta \frac{N^{*}_{n^{}_{r}}-\kappa^{*}}{2\left(n^{}_{r}+2\gamma^{*}\right)}
L_{n^{}_{r}}^{2\gamma^{*}}\left(2\zeta r\right)
\biggr\}.
\end{multline}
Here $\beta \mp 1$ is used to represent large$-$ and small$-$component of the spinor orbitals, respectively. Through application of Eq. (\ref{eq:26}) within the standard derivation for Laguerre polynomials in the form,
\begin{multline}\label{eq:30}
x\frac{d}{dx}L_{q^{*}-p^{*}}^{p^{*}}\left(x\right)
=\left(q^{*}-p^{*}\right)L_{q^{*}-p^{*}}^{p^{*}}\left(x\right)
\\
+\left(q^{*}\right)L_{q^{*}-p^{*}-1}^{p^{*}}\left(x\right)
\end{multline}
and
\begin{multline}\label{eq:31}
x\frac{d}{dx}L_{q^{*}-p^{*}}^{p^{*}}\left(x\right)
=\left(q^{*}-p^{*}+1\right)L_{q^{*}-p^{*}+1}^{p^{*}}\left(x\right)
\\
-\left(q^{*}+1-x\right)L_{q^{*}-p^{*}}^{p^{*}}\left(x\right)
\end{multline}
we obtain the differential equation corresponding to the solution in Eq. (\ref{eq:29}) as,
\begin{multline}\label{eq:32}
\frac{d}{dr}{f^{\varepsilon}}^{\beta}_{n^{}_{r}l^{*}}\left(2\zeta r\right)
=-\beta \frac{\kappa^{*}}{r}
{f^{\varepsilon}}^{\beta}_{n^{}_{r}l^{*}}\left(2\zeta r\right)
\\
+\left(\frac{\beta N^{*}_{n^{}_{r}}-\gamma^{*}-n^{*}_{}}{r} +\zeta\right)
{f^{\varepsilon}}^{-\beta}_{n^{}_{r}l^{*}}\left(2\zeta r\right),
\end{multline}
with,
\begin{align}\label{eq:33}
\begin{array}{ll}
 l\hspace{0.75mm}=\hspace{0.75mm}\begin{cases}
  -\kappa^{*}-\varepsilon & \text{if} \hspace{3mm} \kappa < 0\\
  \kappa^{*}-\varepsilon+1 & \text{if} \hspace{3mm} \kappa > 0
 \end{cases}
\\
l^{*}= \begin{cases}
 -\kappa^{*} & \text{if} \hspace{3mm} \kappa < 0\\
 \kappa^{*}+1 & \text{if} \hspace{3mm} \kappa > 0
 \end{cases}
 \end{array}
\end{align}

The description of integrable systems involves a set of independent commuting operators, given by:
\begin{align}\label{eq:34}
\mathcal{H}=\left\lbrace
{
\hat{H_{1}},
\hat{H_{2}}, 
\hat{H_{3}}, ...,
\hat{H}_{k}
}
\right\rbrace,
\end{align}
$k$ represents the number of degrees of freedom, while the set $\mathcal{H}$ comprises the Hamiltonian and operators of the integrals of motion, namely, invariants. Integrability of Dirac Hamiltonian for an electron in the Coulomb potential is provided by the following sets of operators \cite{43_Eremko_2023},
\begin{align}\label{eq:35}
\mathcal{H}=\left\lbrace
{
\hat{H}_{D},
\hat{J}^{2}, 
\hat{j}_{z},
\hat{K}
}
\right\rbrace,
\end{align}
$\hat{H}_{D}$ the Dirac$-$Hamiltonian given in the Eq. (\ref{eq:2}), $\hat{J}$ is the total angular momentum operator with,
\begin{align}\label{eq:36}
\hat{J}=\
\hat{L}\hat{I}+\dfrac{1}{2}\hat{\Sigma}
\end{align}
and $\hat{j}_{z}$ is projection of $\hat{J}$. $\hat{K}$ is the Dirac invariant:
\begin{align}\label{eq:37}
\hat{K}=\hat{\Pi} . \hat{L}+ \hat{\beta}
\end{align}
\begin{align}\label{eq:38}
\hat{\alpha} &=
\begin{pmatrix}
0 & \hat{\sigma}
\\
\hat{\sigma} & 0
\end{pmatrix},
&
\hat{\beta} &=
\begin{pmatrix}
\hat{I} & 0
\\
0 & -\hat{I}
\end{pmatrix},
&
\hat{\Sigma} &=
\begin{pmatrix}
\hat{\sigma} & 0
\\
0 & \hat{\sigma}
\end{pmatrix},
\end{align}
\begin{align}\label{eq:39}
\hat{\Pi}=\hat{\beta}\hat{\Sigma}=
\begin{pmatrix}
\hat{\sigma} & 0
\\
0 & -\hat{\sigma}
\end{pmatrix},
\end{align} 
$I$ is $2 \times 2$ identy matrix. In accordance with Eq. (\ref{eq:33}), the connection between the eigenvalues of the total angular momentum operator $\hat{J}$ and the operator $\hat{K}$ is determined by,
\begin{align}\label{eq:40}
\kappa^{*} =
\begin{cases}
-\left(j+\varepsilon - \frac{1}{2} \right) & \text{if} \hspace{3mm} \kappa^{*}<0
 \vspace{1mm}\\
+\left(j+\varepsilon - \frac{1}{2} \right) & \text{if} \hspace{3mm} \kappa^{*}>0
 \end{cases}.
\end{align}
Based on Eqs. (\ref{eq:36}, \ref{eq:40}) and the methodology used to derive the general space$-$time \cite{44_Adak_2003, 45_Adak_2004, 46_Martinez_2013, 47_Coletti_2013} formulations for the Dirac and Schr{\"o}dinger equations, it becomes evident that a re$-$definition of the kinetic energy operator is required. Such revision calls for altering the expression for the momentum operator. Derivation of a Hermitian momentum operator in atomic units $\left(a.u.\right)$, grounded in the connection between the Dirac$-$Landau identy and the explicit matrix form expression for the total angular momentum operator,
\begin{align}\label{eq:41}
\left(\hat{\sigma} . \hat{A}\right)
\left(\hat{\sigma} . \hat{B}\right)
=
\left(\hat{A} . \hat{B}\right)
+ i . \hat{\sigma} . \hat{A} \times \hat{B},
\end{align}
\begin{align}\label{eq:42}
\hat{J}^{2}=
\begin{pmatrix}
\left(\hat{L}^{2}+\frac{3}{4}\right)\hat{I} + \hat{\sigma}\hat{L} & 0
\\
0 & \left(\hat{L}^{2}+\frac{3}{4}\right)\hat{I} + \hat{\sigma}\hat{L}
\end{pmatrix},
\end{align}
respectively. Building upon the analysis and the consistency criteria established in Eqs. (\ref{eq:32}, \ref{eq:33}, \ref{eq:35}, \ref{eq:36}, \ref{eq:37}), and Eq. (\ref{eq:40}), we postulate the following form for the kinetic energy operator. A comprehensive exploration of its properties and ramifications is beyond the scope of this study and is reserved for next investigation.\\
The $\left(\vec{\sigma} . \hat{\vec{p}}\right)$ is replaced by,
\begin{multline}\label{eq:43}
\Big(\vec{\sigma} . \hat{\vec{p}}_{\varepsilon}\Big)
=\frac{1}{r^2}\Big(\hat{\sigma} . \hat{r}\Big)
\Big(\hat{\sigma} . \hat{r}\Big)
\Big(\hat{\sigma} \times \hat{\vec{p}}_{\varepsilon}\Big)
\\
=\frac{1}{r^2}\Big(\hat{\sigma} . \hat{r}\Big)
\Big[
\hat{r} . \hat{\vec{p}}_{\varepsilon}
+i \left(\hat{\sigma} . \hat{L}_{\varepsilon} \right)
\Big],
\end{multline}
\begin{align}\label{eq:44}
\Big[
\hat{r} . \hat{\vec{p}}_{\varepsilon}
+i \left(\hat{\sigma} . \hat{L}_{\varepsilon} \right)
\Big]
=-ig\left(r, \varepsilon\right)\frac{\partial}{\partial r}
+ i \left(\hat{\sigma} . \hat{L}_{\varepsilon} \right)
\end{align}
Eq. (\ref{eq:32}) provides an assumption for $g\left(r, \varepsilon\right)$ as $g\left(r,\varepsilon\right)=r$. Rewriting Eq. (\ref{eq:42}) in accordance with Eqs. (\ref{eq:43}, \ref{eq:44}),
\begin{multline}\label{eq:45}
\hat{J}_{\varepsilon}^{2}
\\
=\begin{pmatrix}
\left(\hat{L}_{\varepsilon}^{2}+\frac{3}{4}\right)\hat{I} + \hat{\sigma} . \hat{L}_{\varepsilon} & 0
\\
0 & \left(\hat{L}_{\varepsilon}^{2}+\frac{3}{4}\right)\hat{I} + \hat{\sigma} . \hat{L}_{\varepsilon}
\end{pmatrix}
\end{multline}
allows to deduce an eigenvalue equation of $\hat{\sigma}\hat{L}_{\varepsilon}$,
\begin{multline}\label{eq:46}
\left(\hat{\sigma} . \hat{L}_{\varepsilon}\right)\Omega^{\beta}_{j^{*}m^{*}}
=\Bigg[
\left(j-\beta\varepsilon+\beta\right)
\left(j-\beta\varepsilon+\beta+1\right)
\Bigg.
\\
\Bigg.
-\left(j-\beta\varepsilon+\frac{\beta}{2}\right)
\left(j-\beta\varepsilon+\frac{\beta}{2}+1\right)
-\frac{3}{4}
\Bigg].
\end{multline}
Eq. (\ref{eq:37}) takes form that,
\begin{align}\label{eq:47}
\hat{K}_{\varepsilon}=\hat{\Pi} . \hat{L} + \varepsilon \hat{\beta}.
\end{align}
The relationship between Dirac invariant and its new fractional form obtained as,
\begin{align}\label{eq:48}
\hat{K}=\hat{K}_{\varepsilon}-\left(1-\varepsilon\right)\hat{\beta}.
\end{align}
Thus, eigenvalues of total $\hat{J}_{\varepsilon}^{2}$ and orbital angular momentum $\hat{L}_{\varepsilon}^{2}$ operators are given as,
\begin{align}\label{eq:49}
\begin{cases}
j^{*}\left(j^{*}+1\right) & \text{for} \hspace{3mm} \hat{J}_{\varepsilon}^{2}
 \vspace{1mm}\\
l^{*}\left(l^{*}+1\right) & \text{for} \hspace{3mm} \hat{L}_{\varepsilon}^{2}
 \end{cases},
\end{align} 
here, $j^{*}=\left(j-\beta\varepsilon+\beta \right)$, $l^{*}=\left(j-\beta\varepsilon+\frac{\beta}{2}\right)=j^{*}-\frac{\beta}{2}$, respectively. Finally, Eq. (\ref{eq:46}) simplifies to,
\begin{multline}\label{eq:50}
\left(\hat{\sigma} . \hat{L}_{\varepsilon}\right)\Omega^{\beta}_{j^{*}m^{*}}\left(\theta, \varphi\right)
\\
=\Bigg[
j^{*}\left(j^{*}+1\right)
-l^{*}\left(l^{*}+1\right)
-\frac{3}{4}
\Bigg] \Omega^{\beta}_{j^{*}m^{*}}\left(\theta, \varphi\right).
\end{multline}
The angular parts $\Omega^{\beta}_{j^{*}m^{*}}\left(\theta, \varphi\right)$ of exponential$-$type spinor orbitals are thefore expressed in terms of \textit{generalized spherical harmonics} defined by Infeld and Hull \cite{48_Infeld_1951},
\begin{multline}\label{eq:51}
\frac{d^{2}}{d\theta^{2}}Y_{l^{*}m^{*}}\left(\theta, \varphi\right)
-\frac{\left(m^{*}+\varepsilon^{\prime}\right)\left(m^{*}+\varepsilon^{\prime}-1\right)}{sin^{2}\theta}Y_{l^{*}m^{*}}\left(\theta, \varphi\right)
\\
+\left(\lambda+{\varepsilon^{\prime}}^{2}\right)Y_{l^{*}m^{*}}\left(\theta, \varphi\right)=0,
\end{multline}
with, $2\varepsilon^{\prime}=2-\varepsilon$, $0\leq \varepsilon \leq 1$. For $\varepsilon=1$, the solution of the Eq. (\ref{eq:51}) reduces to standard spherical harmonics $Y_{lm}\left(\theta, \varphi\right)$.
\section{Application of the Bi$-$directional method to compute higher transcendental functions in the Dirac equation solution}
This section is devoted to establish necessary mathematical foundation for solving the Dirac or Dirac$-$like equation, aiming to provide a comprehensive understanding of the mathematical framework required for solution of such equations.
\subsection{Revisiting the Bi$-$directional Hyper$-$radial functions}
Radial parts of one$-$center two$-$electron Coulomb energy \cite{49_Pitzer_1982} in a set of Slater$-$type orbitals \cite{50_Slater_1930, 51_Zener_1930, 52_Slater_1930, 53_Parr_1957},
\begin{align}
\chi_{n^{*}lm}\left(\vec{r}, \zeta \right)
=r^{n^{*}-1}e^{-\zeta r}Y_{lm}\left(\theta,\varphi\right),
\end{align}
\begin{multline}\label{eq:53}
R^{L}_{n^{*}_{1},{n^{\prime}}^{*}_{1},n^{*}_{2},{n^{\prime}}^{*}_{2}}\left(\zeta^{}_{1},\zeta^{\prime}_{1},\zeta^{}_{2},\zeta^{\prime}_{2}\right)
\\
=\int_{0}^{\infty}\int_{0}^{\infty}
r_{1}^{n^{*}_{1}+{n^{\prime}}^{*}_{1}}e^{-\left(\zeta^{}_{1}+\zeta^{\prime}_{1}\right){r_{1}}}
\left(\dfrac{r_{<}^{L}}{r_{>}^{L+1}}\right)
\\
\times r_{2}^{n^{*}_{2}+{n^{\prime}}^{*}_{2}}e^{-\left(\zeta^{}_{2}+\zeta^{\prime}_{2}\right){r_{2}}}
dr_{1}dr_{2},
\end{multline}
are represented in terms of Gauss hyper$-$geometric functions through the Laplace expansion for Coulomb interaction \cite{21_Bagci_2024, 54_Allouche_1974} as,
\begin{multline}\label{eq:54}
R^{L}_{n^{*},{n^{\prime}}^{*}}\left(\zeta,\zeta^{\prime}\right)
=\dfrac{\Gamma\left(n^{*}+{n^{\prime}}^{*}+1\right)}{\left(\zeta+\zeta^{\prime}\right)^{n^{*}+{n^{\prime}}^{*}+1}}
\Bigg\{
\dfrac{1}{n^{*}+L+1}
\\
\times {_2}F_{1}\left[1,n^{*}+{n^{\prime}}^{*}+1,n^{*}+L+2;\frac{\zeta}{\zeta+\zeta^{\prime}}\right]
+\dfrac{1}{{n^{\prime}}^{*}+L+1}
\\
\times {_2}F_{1}\left[1,n^{*}+{n^{\prime}}^{*}+1,{n^{\prime}}^{*}+L+2;\frac{\zeta^{\prime}}{\zeta+\zeta^{\prime}}\right]
\Bigg\},
\end{multline}
here, $n^{*}_{1}+{n^{\prime}}^{*}_{1}$ and $n^{*}_{2}+{n^{\prime}}^{*}_{2}$ is replaced by $n^{*}$ and ${n^{\prime}}^{*}$, similarly $\zeta_{1}+\zeta^{\prime}_{1}$ and $\zeta_{2}+\zeta^{\prime}_{2}$ by $\zeta$ and $\zeta^{\prime}$, respectively. 
${}_{2}F_{1}\left(a,b;c;x\right)$ are defined as the series for $\vert x \vert <1$ \cite{20_Abramowitz_1974, 37_Magnus_1966},
\begin{align}\label{eq:55}
{_2}F_{1}\left[a,b; c; x\right] =\sum_{k=0}^{\infty} \frac{\left(a\right)_{k}\left(b\right)_{k}}{\left(c\right)_{k}} \frac{1}{k!} x^{k}.   
\end{align}
Computing Gauss hyper$-$geometric functions is challenging due to the slow or non$-$convergent nature of their series representations, numerical instability for values near the boundaries of feasibility, and the need for specialized algorithms to ensure precision and efficiency. The functions diverse special cases also demand tailored treatment and computational methodologies, contributing to the complexity of its computation \cite{55_Pearson_2017}. One of the authors has recently chosen to depart from purely mathematical methods for computing the hyper$-$geometric functions, opting instead to leverage the symmetry of Coulomb energy. This approach not only emancipates the Coulomb energy from these functions but also expresses them in terms of Coulomb energy. This approximation referred to as bi$-$directional, has resulted in identification of novel hyper$-$radial functions \cite{21_Bagci_2024}
\begin{multline}\label{eq:56}
{^{+1}}\mathfrak{R}^{L}_{n^{*},{n^{\prime}}^{*}}\left(\zeta,\zeta^{\prime}\right)
\\
=\dfrac{R^{L}_{n^{*},{n^{\prime}}^{*}}\left(\zeta,\zeta^{\prime}\right)+{^{+1}}m^{L}_{{n^{\prime}}^{*} n^{*}}\left(\zeta^{\prime}, \zeta\right)}{e^{L}_{n^{*} {n^{\prime}}^{*}} \hspace{1mm} {^{+1}}h_{{n^{\prime}}^{*} n^{*}}^{L}\left(\zeta^{\prime},\zeta\right)}.
\end{multline}
\begin{multline}\label{eq:57}
{_2}F_{1}\left[1,n^{*}+{n^{\prime}}^{*}+1,n^{*}+L+2;\frac{\zeta}{\zeta+\zeta^{\prime}}\right]
\\
=\dfrac{R^{L}_{n^{*},{n^{\prime}}^{*}}\left(\zeta, \zeta^{\prime}\right)+{^{+1}}m^{L}_{{n^{\prime}}^{*} n^{*}}\left(\zeta^{\prime},\zeta\right)}{e^{L}_{n^{*} {n^{\prime}}^{*}}{^{+1}}h_{{n^{\prime}}^{*} n^{*}}^{L}\left(\zeta^{\prime},\zeta\right)},
\end{multline}
\begin{multline}\label{eq:58}
{_2}F_{1}\left[1,n^{*}+{n^{\prime}}^{*}+1,{n^{\prime}}^{*}+L+2;\frac{\zeta^{\prime}}{\zeta+\zeta^{\prime}}\right]
\\
=\dfrac{R^{L}_{{n^{\prime}}^{*},n^{*}}\left(\zeta^{\prime}, \zeta\right)+{^{+1}}m^{L}_{n^{*} {n^{\prime}}^{*}}\left(\zeta, \zeta^{\prime}\right)}{e^{L}_{{n^{\prime}}^{*} n^{*}}{^{+1}}h_{n^{*} {n^{\prime}}^{*}}^{L}\left(\zeta, \zeta^{\prime}\right)},
\end{multline}
${^{+1}}\mathfrak{R}^{L}_{n^{*},{n^{\prime}}^{*}}$ are the hyper$-$radial functions. The auxiliary functions $e^{L}_{n^{*} {n^{\prime}}^{*}}$, ${^{+1}}h_{{n^{\prime}}^{*} n^{*}}^{L}$ arising in Eq. (\ref{eq:56}) are given as,
\begin{align}\label{eq:59}
e^{L}_{n^{*} {n^{\prime}}^{*}}
=\dfrac{\Gamma\left(n^{*}+{n^{\prime}}^{*}+1\right)}{\left(\zeta+\zeta^{\prime}\right)^{n^{*}+{n^{\prime}}^{*}+1}}\dfrac{1}{n^{*}+L+1},
\end{align}
\begin{multline}\label{eq:60}
{^{+1}}h_{n^{*} {n^{\prime}}^{*}}^{L}\left(\zeta, \zeta^{\prime}\right)
=\dfrac{{n^{\prime}}^{*}+L+1}{n^{*}+L+1}\dfrac{\left(n^{*}-L\right)_{2L+1}}{\left(-{n^{\prime}}^{*}-L-1\right)_{2L+1}}
\\
\times \left(-\dfrac{\zeta^{\prime}}{\zeta}\right)^{2L+1}
f_{n^{*} {n^{\prime}}^{*}}^{L}+1,
\end{multline}
with,
\begin{align}\label{eq:61}
f_{n^{*} {n^{\prime}}^{*}}^{L}=
\dfrac{\pi csc\left[\left(-{n^{\prime}}^{*}+L\right)\pi\right]}{\Gamma\left(-{n^{\prime}}^{*}+L+1\right)}\dfrac{\left(n^{*}+L+1\right)}{\Gamma\left({n^{\prime}}^{*}-L+1\right)}.
\end{align}
The ${^{+1}}m^{L}_{{n^{\prime}}^{*},n^{*}}$ auxiliary functions are determined by,
\begin{multline}\label{eq:62}
{^{+1}}m^{L}_{n^{*} {n^{\prime}}^{*}}\left(\zeta, \zeta^{\prime}\right)
\\
=e^{L}_{n^{*} {n^{\prime}}^{*}}
\left\{g_{n^{*} {n^{\prime}}^{*}}^{L}\left(\zeta,\zeta^{\prime}\right)+{^{+1}}l_{n^{*} {n^{\prime}}^{*}}^{L}\left(\zeta,\zeta^{\prime}\right)
\right\},
\end{multline}
with,
\begin{multline}\label{eq:63}
g_{n^{*} {n^{\prime}}^{*}}^{L}\left(\zeta,\zeta^{\prime}\right)
=f_{n^{*} {n^{\prime}}^{*}}\dfrac{\Gamma\left({n^{\prime}}^{*}-L+1\right)\Gamma\left(n^{*}+L+1\right)}{\Gamma\left(n^{*}+{n^{\prime}}^{*}+1\right)}
\\
\times \left(\frac{\zeta}{\zeta+\zeta^{\prime}}\right)^{-n^{*}-L-1}
\left(\frac{\zeta^{\prime}}{\zeta+\zeta^{\prime}}\right)^{-{n^{\prime}}^{*}+L},
\end{multline}
and,
\begin{multline}\label{eq:64}
{^{+1}}l_{n^{*} {n^{\prime}}^{*}}^{L}\left(\zeta, \zeta^{\prime}\right)
=\dfrac{\pi csc\left[\left(-{n^{\prime}}^{*}+L\right)\pi\right]}{\Gamma\left(-{n^{\prime}}^{*}+L+1\right)}
\dfrac{\left(n^{*}+L+1\right)}{\Gamma\left({n^{\prime}}^{*}-L+1\right)}
\\
\times \left(-\dfrac{\zeta^{\prime}}{\zeta}\right)^{2L+1}
\left(\dfrac{\zeta+\zeta^{\prime}}{\zeta^{\prime}}\right)
\\
\times \sum_{k=1}^{2L+1}\dfrac{\left(n^{*}-L+k\right)_{2L+1-k}}{\left(-{n^{\prime}}^{*}-L-1+k\right)_{2L+1-k}}\left(-\dfrac{\zeta}{\zeta^{\prime}}\right)^{k-1}.
\end{multline}
Notice that, explicit computation of the hyper$-$radial functions via Eq. (\ref{eq:56}) offers no essential advantage in calculation of Coulomb energy. It can readily be obtained using the provided recurrence relations in \cite{21_Bagci_2024}:
\begin{multline}\label{eq:65}
{^{+1}}\mathfrak{R}^{L+2}_{n^{*},{n^{\prime}}^{*}}\left(\zeta,\zeta^{\prime}\right)
=\dfrac{\left(n^{*}+L+3\right)}{\zeta\left(n^{*}+L+2\right)\left(-{n^{\prime}}^{*}+L+2\right)}
\\
\times \bigg\{
\zeta^{\prime} \left(n^{*}+L+2\right){^{+1}}\mathfrak{R}^{L}_{n^{*},{n^{\prime}}^{*}}\left(\zeta,\zeta^{\prime}\right)
\\
+\left[\zeta \left(-{n^{\prime}}^{*}+L+1\right)-\zeta^{\prime}\left(n^{*}+L+2\right)\right]
\\
\times
{^{+1}}\mathfrak{R}^{L+1}_{n^{*},{n^{\prime}}^{*}}\left(\zeta,\zeta^{\prime}\right)
\bigg\},
\end{multline}
here, for $L=0$,
\begin{multline}\label{eq:66}
{^{+1}}\mathfrak{R}^{0}_{n^{*},{n^{\prime}}^{*}}\left(\zeta,\zeta^{\prime}\right)=
\dfrac{R^{0}_{n^{*},{n^{\prime}}^{*}}\left(\zeta,\zeta^{\prime}\right)+{^{+1}}m^{0}_{{n^{\prime}}^{*} n^{*}}\left(\zeta^{\prime},\zeta\right)}{e^{0}_{n^{*} {n^{\prime}}^{*}} \hspace{1mm} {^{+1}}h_{{n^{\prime}}^{*} n^{*}}^{0}\left(\zeta^{\prime}, \zeta\right)}
\\
=\left(n^{*}+1\right)\left(\dfrac{\zeta}{\zeta+\zeta^{\prime}}\right)^{-n^{*}-1}
\left(\dfrac{\zeta^{\prime}}{\zeta+\zeta^{\prime}}\right)^{-{n^{\prime}}^{*}}
\\
\times \Bigg\{
\dfrac{\Gamma\left(n^{*}+1\right)\Gamma\left({n^{\prime}}^{*}\right)}{\Gamma\left(n^{*}+{n^{\prime}}^{*}+1\right)}
-B_{{n^{\prime}}^{*},n^{*}+1}\left(\dfrac{\zeta^{\prime}}{\zeta+\zeta^{\prime}}\right)
\Bigg\},
\end{multline}
with $B_{n^{*} {n^{\prime}}^{*}}$ are the incomplete beta functions \cite{20_Abramowitz_1974}. For $L=1$,
\begin{multline}\label{eq:67}
{^{+1}}\mathfrak{R}^{1}_{n^{*},{n^{\prime}}^{*}}\left(\zeta,\zeta^{\prime}\right)
\\
=\left(\dfrac{n^{*}+2}{n^{\prime}-1}\right)
\bigg[
\left(\dfrac{\zeta^{\prime}}{\zeta}\right)
{^{+1}}\mathfrak{R}^{0}_{{n^{\prime}}^{*},n^{*}}\left(\zeta,\zeta^{\prime}\right)
-\left(\dfrac{\zeta+\zeta^{\prime}}{\zeta}\right)
\bigg].
\end{multline}
The above relationships obtained for Coulomb energy involve hypergeometric functions with a parameter $c=n+L+2$. The formulas presented in \cite{21_Bagci_2024} allow for an alternative representation of hyper$-$radial functions, where the Coulomb energy this time involves hypergeometric functions with a parameter $c=n^{*}-L+1$. Given these conditions, the hyper-radial functions would be more appropriately formulated as follows,
\begin{align}\label{eq:68}
\mathfrak{R}^{L}_{n^{*},{n^{\prime}}^{*}}\left(\zeta,\zeta^{\prime}\right)
=
\begin{bmatrix}
{^{+1}}\mathfrak{R}^{L}_{n^{*},{n^{\prime}}^{*}}\left(\zeta,\zeta^{\prime}\right)
\\
{^{-1}}\mathfrak{R}^{L}_{n^{*},{n^{\prime}}^{*}}\left(\zeta,\zeta^{\prime}\right)
\end{bmatrix},
\end{align}
\begin{align}\label{eq:69}
{^{\beta}}\mathfrak{R}^{L}_{n^{*},{n^{\prime}}^{*}}\left(\zeta,\zeta^{\prime}\right)
=\dfrac{R^{L}_{n^{*},{n^{\prime}}^{*}}\left(\zeta,\zeta^{\prime}\right)+{^{\beta}}m^{L}_{{n^{\prime}}^{*} n^{*}}\left(\zeta^{\prime}, \zeta\right)}{e^{L}_{n^{*} {n^{\prime}}^{*}} \hspace{1mm} {^{\beta}}h_{{n^{\prime}}^{*} n^{*}}^{L}\left(\zeta^{\prime},\zeta\right)}.
\end{align}
\begin{multline}\label{eq:70}
{^{\beta}}m^{L}_{n^{*} {n^{\prime}}^{*}}\left(\zeta, \zeta^{\prime}\right)
\\
=e^{L}_{n^{*} {n^{\prime}}^{*}}
\left\{g_{n^{*} {n^{\prime}}^{*}}^{L}\left(\zeta,\zeta^{\prime}\right)+\beta \hspace{1mm} {^{\beta}}l_{n^{*} {n^{\prime}}^{*}}^{L}\left(\zeta,\zeta^{\prime}\right)
\right\}.
\end{multline}
For the specific expressions of ${^{\beta}}l_{n^{*} {n^{\prime}}^{*}}^{L}$ and ${^{\beta}}h_{{n^{\prime}}^{*} n^{*}}^{L}$ with $\beta=-1$, please see \cite{21_Bagci_2024}. The following connections between two form of hyper$-$radial functions are obtained,
\begin{multline}\label{eq:71}
{e^{L}_{n^{*} {n^{\prime}}^{*}} \hspace{1mm} {^{+1}}h_{{n^{\prime}}^{*} n^{*}}^{L}\left(\zeta^{\prime},\zeta\right)} \hspace{1mm} {^{+1}}\mathfrak{R}^{L}_{n^{*},{n^{\prime}}^{*}}\left(\zeta,\zeta^{\prime}\right)
\\
+
{e^{L}_{n^{*} {n^{\prime}}^{*}} \hspace{1mm} {^{-1}}h_{{n^{\prime}}^{*} n^{*}}^{L}\left(\zeta^{\prime},\zeta\right)} \hspace{1mm} {^{-1}}\mathfrak{R}^{L}_{n^{*},{n^{\prime}}^{*}}\left(\zeta,\zeta^{\prime}\right)
\\
={^{+1}}m^{L}_{n^{*} {n^{\prime}}^{*}}\left(\zeta, \zeta^{\prime}\right)
+
{^{-1}}m^{L}_{n^{*} {n^{\prime}}^{*}}\left(\zeta, \zeta^{\prime}\right),
\end{multline}
\begin{align}\label{eq:72}
{^{+1}}\mathfrak{R}^{L}_{n^{*},{n^{\prime}}^{*}}\left(\zeta,\zeta^{\prime}\right)
={^{-1}}\mathfrak{R}^{L}_{n^{*}+2L+1,{n^{\prime}}^{*}-\left(2L+1\right)}\left(\zeta,\zeta^{\prime}\right).
\end{align}
The recurrence relationships for ${^{-1}}\mathfrak{R}^{L}_{n^{*},{n^{\prime}}^{*}}$ are derived as,
\begin{multline}\label{eq:73}
{^{-1}}\mathfrak{R}^{-\left(L+2\right)}_{n^{*},{n^{\prime}}^{*}}\left(\zeta,\zeta^{\prime}\right)
=\dfrac{\left(n^{*}-L+2\right)}{\zeta\left(n^{*}-L+1\right)\left(-{n^{\prime}}^{*}-L+1\right)}
\\
\times \bigg\{
\zeta^{\prime} \left(n^{*}-L+1\right){^{-1}}\mathfrak{R}^{-L}_{n^{*},{n^{\prime}}^{*}}\left(\zeta,\zeta^{\prime}\right)
\\
+\left[\zeta \left(-{n^{\prime}}^{*}-L\right)-\zeta^{\prime}\left(n^{*}-L+1\right)\right]
\\
\times
{^{-1}}\mathfrak{R}^{-\left(L+1\right)}_{n^{*},{n^{\prime}}^{*}}\left(\zeta,\zeta^{\prime}\right)
\bigg\},
\end{multline}
\begin{multline}\label{eq:74}
{^{-1}}\mathfrak{R}^{0}_{n^{*},{n^{\prime}}^{*}}\left(\zeta,\zeta^{\prime}\right)=
\dfrac{R^{0}_{n^{*},{n^{\prime}}^{*}}\left(\zeta,\zeta^{\prime}\right)+{^{-1}}m^{0}_{{n^{\prime}}^{*} n^{*}}\left(\zeta^{\prime},\zeta\right)}{e^{0}_{n^{*} {n^{\prime}}^{*}} \hspace{1mm} {^{-1}}h_{{n^{\prime}}^{*} n^{*}}^{0}\left(\zeta^{\prime}, \zeta\right)}
\\
=n\left(\dfrac{\zeta}{\zeta+\zeta^{\prime}}\right)^{-n^{*}}
\left(\dfrac{\zeta^{\prime}}{\zeta+\zeta^{\prime}}\right)^{-{n^{\prime}}^{*}-2}
\\
\times \Bigg\{
\dfrac{\Gamma\left(n^{*}\right)\Gamma\left({n^{\prime}}^{*}+2\right)}{\Gamma\left(n^{*}+{n^{\prime}}^{*}+2\right)}
-B_{{n^{\prime}}^{*}+2,n^{*}}\left(\dfrac{\zeta^{\prime}}{\zeta+\zeta^{\prime}}\right)
\Bigg\},
\end{multline}
\begin{multline}\label{eq:75}
{^{-1}}\mathfrak{R}^{-1}_{n^{*},{n^{\prime}}^{*}}\left(\zeta,\zeta^{\prime}\right)
=\left(\dfrac{n^{*}}{{n^{\prime}}^{*}+1}\right)
\\
\times \bigg[
\left(\dfrac{\zeta^{\prime}}{\zeta}\right)
{^{-1}}\mathfrak{R}^{0}_{{n^{\prime}}^{*},n^{*}}\left(\zeta,\zeta^{\prime}\right)
-\left(\dfrac{\zeta+\zeta^{\prime}}{\zeta}\right)
\bigg].
\end{multline}
\subsection{Computational Aspect for ${^{\mp 1}}\mathfrak{R}^{0}_{n,n^{\prime}}$}
The hyper$-$radial functions resulting from the aforementioned manipulations are a direct consequence of the Laplace expansion of the Coulomb potential. It is an example of two$-$range addition theorem defined in three$-$dimensional Taylor expansion \cite{36_Weniger_2008, 56_Weniger_2000}, exhibiting point$-$wise convergence. The Laplace expansion has a two$-$range form, depending on the relative length of $r_{1}$ and $r_{2}$. Despite its purely mathematical nature, the accurate computation of right$-$hand side of Eq. (\ref{eq:69}) is intricately linked to the formulation of the Coulomb interaction. Conversely, the hyper-radial functions determines the accuracy of the Coulomb energy. Eqs. (\ref{eq:66}) and (\ref{eq:75}) provide further examples for minimum value of $L$, $L=0$, as they involve incomplete beta functions. New series representations for these functions can be obtained through the expansion of the Coulomb interaction using various methods relying on diverse single$-$center expansions converging this time in the mean.\\
Building upon the radial integral derived in Eq. (\ref{eq:53}) and ensuring subsequent analysis adheres rigorously to the principles of the Laplace expansion: the relative dependence on $r_1$ and $r_2$ may be removed as \cite{57_Fontana_1961},
\begin{multline}\label{eq:76}
\frac{r_{<}^{L}}{r_{>}^{L+1}}
\\
=\left(2L+1\right)
\sum_{k=L}^{\infty}{}^{\prime}
\frac{r_{1}^{k} r_{2}^{k}\left(2k+1\right)!!}{\left(k+L+1\right)!!\left(k-L\right)!! r^{2k+1}},
\end{multline}
here, the prime notation in the summation denotes a step$-$wise increment of $2$, and $\left(a\right)!!$ represents double factorial of $a$. $r=\sqrt{r_{1}+r_{2}}$. Setting $L=0$, after a bit of manipulations, using Eq. (\ref{eq:76}) in Eq. (\ref{eq:53}) results in a convergent series representation for incomplete beta functions.\\
If we consider to start from the inter$-$electronic separation of $\vec{r}_{12}$,
\begin{align}\label{eq:77}
\left(\vec{r}_{12}\right)^2
=\left(\vec{r}_{1}+\vec{r}_{2}\right)^{2}-
2\vec{r}_{1}\vec{r}_{2}\left(1+cos\theta\right),
\end{align}
using the binomial theorem and Laplace integral transform, the Coulomb interaction is given as \citep{58_Roberts_1966},
\begin{multline}\label{eq:78}
\frac{1}{\vert \vec{r}_{1}-\vec{r}_{2} \vert}
=\sum_{k=0}^{\infty} \frac{\left(1+cos\theta\right)^{k}}{2^{k}\left(k!\right)^{2}}\left(r_{1} r_{2}\right)^{k}
\\
\times
\int_{0}^{\infty} u^{2k}e^{-u r_{1}-u r_{2}}du.
\end{multline}
Substituting Eq. (\ref{eq:78}) into the Coulomb energy expression yields a radial integral that is represented by an infinite series deviates from Eq. (\ref{eq:53}) only by angular coefficients,
\begin{multline}\label{eq:79}
R^{L}_{n^{*}_{1},{n^{\prime}}^{*}_{1},n^{*}_{2},{n^{\prime}}^{*}_{2}}\left(\zeta^{}_{1},\zeta^{\prime}_{1},\zeta^{}_{2},\zeta^{\prime}_{2}\right)
\\
=V^{L}_{l_{1}m_{1},l^{\prime}_{1}m^{\prime}_{1}; l_{2}m_{2},l^{\prime}_{2}m^{\prime}_{2}}
\hspace{1mm}
{^{o}}R_{n^{*}_{1},{n^{\prime}}^{*}_{1},n^{*}_{2},{n^{\prime}}^{*}_{2}}\left(\zeta^{}_{1},\zeta^{\prime}_{1},\zeta^{}_{2},\zeta^{\prime}_{2}\right).
\end{multline}
Proceeding with the one$-$range addition theorem \cite{36_Weniger_2008, 59_Steinborn_1980, 60_Steinborn_1982}, the Coulomb interaction is expanded in terms of complete and orthonormal exponential$-$type orbitals within the Hilbert space $\left[L^{2}\left(\mathbb{R}^{3}\right)\right]$. Notice that extending the formalism of one$-$range addition theorem to functions do not belong to the Hilbert space may lead to divergence with respect to their corresponding norm, termed \textit{weak convergence} \cite{36_Weniger_2008}. Exponential$-$type orbitals with non$-$integer quantum numbers given in [Eq. (\ref{eq:12})]  for $\alpha=0$  are complete and orthonormal with respect to $L^{2}\left(\mathbb{R}^{3}\right)$. Dropping the restriction of quantum number for hydrogenic orbitals provide further advantages in expansion of Coulomb interaction,
\begin{align}\label{eq:80}
\frac{1}{\vert \vec{r}_{1}-\vec{r}_{2} \vert}
=\sum_{n^{*}lm}A_{n^{* \prime}l^{\prime}m^{\prime}}\left(\vec{r_{2}}, \zeta\right)
\Psi_{n^{*}lm}^{0 \varepsilon}\left(\vec{r_{1}},\zeta\right),
\end{align}
where,
\begin{multline}\label{eq:81}
A_{n^{* \prime}l^{\prime}m^{\prime}}\left(\vec{r_{2}}, \zeta\right)
\\
=\int_{0}^{\infty} 
\frac{1}{\vert \vec{r}_{1}-\vec{r}_{2} \vert}
\left[\Psi_{n^{* \prime}l^{\prime}m^{\prime}}^{0 \varepsilon}\left(\vec{r_{1}},\zeta\right)\right]^{*} d\vec{r}_{1}.
\end{multline}
Given the clear separation of variables $\vec{r}_{1}$, $\vec{r}_{2}$ and ensured convergence, this method is also applicable for the computation of incomplete beta functions while $L=0$. Computing incomplete beta functions arising in Eqs. (\ref{eq:66}, \ref{eq:75}) through direct use of Eq. (\ref{eq:53}), without relying solely on mathematical approximations and capitalizing on the symmetry of the Coulomb interaction implies \textit{self$-$consistency}. This observation reinforces the notion that the symmetries inherent in physical systems provide essential tools for overcoming potential mathematical challenges within them.\\
Simplification and faster convergence may be possible for large values of parameters by expanding the power functions in Eq. (\ref{eq:53}) in terms of $\Psi_{n^{*}lm}^{\alpha \varepsilon}$ as \cite{14_Bagci_2023},
\begin{align}\label{eq:82}
\chi_{n^{*}l^{*}m^{*}}\left(\zeta,\vec{r}\right)
=\sum_{{n'}^{*}=l^{*}+\epsilon}^{n^{*}}
\tilde{a}^{\alpha \epsilon l^{*}}_{n^{*}{n'}^{*}}\Psi^{\alpha \epsilon}_{{n'}^{*}l^{*}m^{*}}\left(\zeta,\vec{r}\right),
\end{align}
here,
\begin{multline}\label{eq:83}
\tilde{a}^{\alpha\epsilon l^{*}}_{n^{*}{n'}^{*}}
=\left(-1\right)^{{n'}^{*}-l^{*}-\epsilon}
\\
\times \Bigg[\dfrac{\left(2{n'}^{*}\right)^\alpha \Gamma\left( n^{*}+l^{*}+\epsilon+1-\alpha \right)}{\Gamma \left( n^{*}+l^{*}+\epsilon+1\right)F_{n^{*}-l^{*}-\epsilon}\left(2n^{*}\right)}
\Bigg.\\
\times F_{{n'}^{*}+l^{*}+\epsilon-\alpha}\left(n^{*}+l^{*}+\epsilon-\alpha\right) \Bigg.
\\
\times F_{{n'}^{*}-l^{*}-\epsilon}\left(n^{*}-l^{*}-\epsilon\right)
\Bigg]^{1/2}.
\end{multline}
\subsection{Asymptotic Analysis for ${^{\mp 1}}\mathfrak{R}^{L}_{n,n^{\prime}}$}
The integrals associated with hyper$-$geometric functions \cite{61_Bateman_1954} are written in a form that:
\begin{multline}\label{eq:84}
\dfrac{1}{\left(\mu+\nu-b\right)^{\nu}}\int_{0}^{\infty}x^{\mu-1}e^{-bx}\Gamma\left[\nu,\left(\mu+\nu-b\right)x\right]dx
\\
=\dfrac{\Gamma\left(\mu+\nu\right)}{\mu\left(\mu+\nu\right)^{\mu+\nu}}
{_2}F_{1}\left[1,\mu+\nu,\mu+1,\frac{b}{\mu+\nu}\right],
\end{multline}
\begin{multline}\label{eq:85}
\dfrac{1}{a^{\nu}}\int_{0}^{\infty}x^{\mu-1}e^{-\left(\mu+\nu-a\right)x}\gamma\left[\nu,ax\right]dx
\\
=\dfrac{\Gamma\left(\mu+\nu\right)}{\nu\left(\mu+\nu\right)^{\mu+\nu}}
{_2}F_{1}\left[1,\mu+\nu,\nu+1,\frac{a}{\mu+\nu}\right].
\end{multline}
The values of $m$ and $n$, extracted from the Laplace expansion of the Coulomb interaction, leading to Eq. (\ref{eq:54}) upon application to Eq. (\ref{eq:53}),
\begin{align}\label{eq:86}
\begin{array}{l}
\nu=n^{*}-L \quad \mu={n^{\prime}}^{*}+L+1, \quad  \text{for} \hspace{3mm} \text{Eq.} (\ref{eq:84})
 \vspace{1mm}\\
\nu=n^{*}+L+1 \quad \mu={n^{\prime}}^{*}-L, \quad \text{for} \hspace{3mm} \text{Eq.} (\ref{eq:85})
 \end{array}.
\end{align}
The hyper$-$geometric functions arising in these equations take form that,
\begin{multline*}
\begin{array}{l}
\text{from Eq.} (\ref{eq:84}),
\\
\rightarrow {_2}F_{1}\left[1,n^{*}+{n^{\prime}}^{*}+1,{n^{\prime}}^{*}+L+2,\dfrac{\zeta^{\prime}}{n^{*}+{n^{\prime}}^{*}+1-\zeta^{\prime}}\right]
\vspace{3mm}\\
\text{from Eq.} (\ref{eq:85}),
\\
 \rightarrow {_2}F_{1}\left[1,n^{*}+{n^{\prime}}^{*}+1,n+L+2,\dfrac{\zeta}{n^{*}+{n^{\prime}}^{*}+1-\zeta}\right]
 \end{array}
\end{multline*}
Using the following connections given between incomplete gamma and confluent hyper$-$geometric functions \cite{62_Arfken_1985},
\begin{align}\label{eq:87}
{_1}F_{1}\left[a,c,x\right]=\lim_{n^{\prime} \rightarrow \infty}
{_2}F_{1}\left[a,b,c,b^{-1}x\right],
\end{align}
and,
\begin{align}\label{eq:88}
{_1}F_{1}\left[1,c+1,x\right]
=\frac{\gamma\left[c,x\right]}{c^{-1}x^{c}e^{-x}}
\end{align}
The relationships between hyper$-$radial functions and incomplete gamma functions are obtained as,
\begin{multline}\label{eq:89}
\gamma\left[{n^{\prime}}^{*}+L+1, \zeta^{\prime}\right]
=\lim_{n^{*} \rightarrow \infty}
\dfrac{{\zeta^{\prime}}^{{n^{\prime}}^{*}+L+1}e^{-\zeta^{\prime}}}{\left({n^{\prime}}^{*}+L+1\right)}
\\
\times {^{+1}}\mathfrak{R}^{L}_{n^{*},{n^{\prime}}^{*}}\left(\zeta^{\prime},n^{*}+{n^{\prime}}^{*}+1-\zeta^{\prime}\right)
\end{multline}
\begin{multline}\label{eq:90}
\gamma\left[n^{*}+L+1, \zeta\right]
=\lim_{{n^{\prime}}^{*} \rightarrow \infty}
\dfrac{\zeta^{n^{*}+L+1}e^{-\zeta}}{\left(n^{*}+L+1\right)}
\\
\times {^{+1}}\mathfrak{R}^{L}_{n^{*},{n^{\prime}}^{*}}\left(\zeta,n^{*}+{n^{\prime}}^{*}+1-\zeta\right).
\end{multline}
Exchanging the values of $\mu$ and $\nu$ in Eq. (\ref{eq:86}) yields analogous formulas for ${^{-1}}\mathfrak{R}^{L}$:
\begin{multline}\label{eq:91}
\gamma\left[{n^{\prime}}^{*}-L, \zeta^{\prime}\right]
=\lim_{n^{*} \rightarrow \infty}
\dfrac{{\zeta^{\prime}}^{{n^{\prime}}^{*}-L}e^{-\zeta^{\prime}}}{\left({n^{\prime}}^{*}-L\right)}
\\
\times {^{-1}}\mathfrak{R}^{L}_{n^{*},{n^{\prime}}^{*}}\left(\zeta^{\prime},n^{*}+{n^{\prime}}^{*}+1-\zeta^{\prime}\right),
\end{multline}
\begin{multline}\label{eq:92}
\gamma\left[n^{*}-L, \zeta\right]
=\lim_{{n^{\prime}}^{*} \rightarrow \infty}
\dfrac{\zeta^{n^{*}-L}e^{-\zeta}}{\left(n^{*}-L\right)}
\\
{^{-1}}\mathfrak{R}^{L}_{n^{*},{n^{\prime}}^{*}}\left(\zeta,n^{*}+{n^{\prime}}^{*}+1-\zeta\right).
\end{multline}

Convergent series expansions for the asymptotics of hyper$-$radial functions are derived. These expansions are expressed in terms of the radial part of exponential type orbitals. The derivation utilizes the following relationships \cite{63_Bateman_1954},
\begin{align}\label{eq:93}
{_1}F_{1}\left[a,c,\frac{x}{x-1}y \right]
=\left(1-x\right)^{a}
\sum_{k=0}^{\infty} \frac{\left(a\right)_{k}}{\left(c\right)_{k}}
L_{k}^{c-1}\left(y\right)x^{k},
\end{align}
where, $\vert x \vert <1$, $y>0$. Substituting $x$ with $x=\frac{\zeta}{\zeta-\zeta^{\prime}}$, $y=2\zeta^{\prime}r$, using Eq. (\ref{eq:88}) for $a=1$ in Eq. (\ref{eq:93}) and re$-$arranging the summation as $k=n-l-\varepsilon$ we obtain,
\begin{multline}\label{eq:94}
\frac{\left(2l+2\varepsilon-\alpha\right)}{\Gamma\left(2l+2\varepsilon-\alpha+1\right)}
\frac{\gamma\left[2l+2\varepsilon-\alpha,2\zeta r\right]}{\left(2\zeta r\right)^{l+\varepsilon-\alpha+1}e^{-\zeta r}}
=\left(\frac{\zeta^{\prime}}{\zeta^{\prime}-\zeta}\right)
\\
\times \lim_{N\rightarrow \infty} \sum_{n^{*}=l+\varepsilon}^{N+l+\varepsilon}
\left(\frac{\zeta}{\zeta-\zeta^{\prime}}\right)^{n^{*}-l-\varepsilon}
\frac{\Gamma\left(n^{*}-l-\varepsilon+1\right)}{\Gamma\left(n^{*}+l+\varepsilon-\alpha+1\right)}
\\
\times
\hspace{1mm}
{^*}R^{\alpha \varepsilon}_{n^{*}l}\left(\zeta, \zeta^{\prime}, r\right).
\end{multline}
The derivative of Eq. (\ref{eq:94}) with respect to $r$ yield an expansion for generalized power functions.
Using the derivative for incomplete gamma functions,
\begin{align}\label{eq:95}
\frac{\partial \gamma\left(a,bx\right)}{\partial x}
=b\left(bx\right)^{a-1}e^{-bx},
\end{align}
we have,
\begin{multline}\label{eq:96}
\frac{\left(2l+2\varepsilon-\alpha\right)}{\Gamma\left(2l+2\varepsilon-\alpha+1\right)}
\left[\left(2\zeta r\right)^{l+\varepsilon-1} e^{-\zeta r} \right]
=\left(\frac{\zeta^{\prime}}{\zeta^{\prime}-\zeta}\right)
\\
\times \lim_{N\rightarrow \infty} \sum_{n^{*}=l+\varepsilon}^{N+l+\varepsilon}
\left(\frac{\zeta}{\zeta-\zeta^{\prime}}\right)^{n^{*}-l-\varepsilon}
\frac{\Gamma\left(n^{*}-l-\varepsilon+1\right)}{\Gamma\left(n^{*}+l+\varepsilon-\alpha+1\right)}
\\
\times
\biggl\{
\left(2\zeta^{\prime} r\right)
\hspace{1mm}
{^*}R^{\alpha \varepsilon}_{n^{*}l}\left(\zeta, \zeta^{\prime}, r\right)
-\left(2\zeta r\right)
\hspace{1mm}
{^*}R^{\alpha \varepsilon}_{n^{*}l}\left(\zeta, \zeta^{\prime}, r\right)
\biggl\}.
\end{multline}
Through the addition formulas for Laguerre functions,
\begin{multline}\label{eq:97}
L_{q^{*}-p^{*}}^{p^{*}}\left(x_{1}x_{2}\right)
=\frac{\Gamma\left(q^{*}+1\right)}{\Gamma\left(q^{*}-p^{*}+k+1\right)\Gamma\left(p^{*}+k+1\right)}
\\
\times
x_{1}^{k}\left(1-x_{1}\right)^{q^{*}-p^{*}-k}L_{k}^{p^{*}}\left(x_{2}\right),
\end{multline}

the radial$-$type exponential functions in Eqs. (\ref{eq:94}, \ref{eq:96}) are expressed as,
\begin{align*}
\zeta^{\prime}=\frac{x-1}{x}\zeta,
\end{align*}
\begin{multline}\label{eq:98}
{^*}R^{\alpha \varepsilon}_{n^{*}l}\left(\zeta, \zeta^{\prime}, r\right)
=\left(2\zeta r\right)^{l+\varepsilon-1} e^{-\zeta r} L_{n^{*}-l-\varepsilon}^{2l+2\varepsilon-\alpha}\left(2\zeta^{\prime}r\right)
\\
=\sum_{{n^{\prime}}^{*}=l+\varepsilon}^{n^{*}}
\left(\frac{\zeta^{\prime}}{\zeta}\right)^{{n^{\prime}}^{*}-l-\varepsilon}
\left(\frac{\zeta-\zeta^{\prime}}{\zeta}\right)^{n^{*}-{n^{\prime}}^{*}}
\\
\times \frac{\Gamma\left(n^{*}+l+\varepsilon-\alpha+1\right)}{\Gamma\left(n^{*}-{n^{\prime}}^{*}+1\right)\Gamma\left(n^{\prime}+l+\varepsilon-\alpha+1\right)}
\\
\times
{^*}R^{\alpha \varepsilon}_{n^{*}l}\left(\zeta, r\right),
\end{multline}
here,
\begin{align}\label{eq:99}
R^{\alpha \varepsilon}_{n^{*}l}\left(\zeta, r\right)
=N_{n^{*}l}^{\alpha \varepsilon}\left(\zeta\right)
\hspace{1mm}
{^*}R^{\alpha \varepsilon}_{n^{*}l}\left(\zeta, r\right),
\end{align}
are the radial parts of complete orthonormal sets of exponential$-$type orbitals with non$-$integer quantum numbers.

\end{document}